# In-situ comparison of interface instability of basal and edge planes during unidirectional growth of sea ice

*Tongxin Zhang, Zhijun Wang\*, Lilin Wang\*, Junjie Li, Jincheng Wang*

*State Key Laboratory of Solidification Processing, Northwestern Polytechnical University, Xi'an 710072, China*

**Abstract:** The unique anisotropy of ice has endowed sea ice growth a peculiar and attractive subject from both fundamental and applied viewpoints. The distinct growth behaviors between edge and basal plane of ice are one of the central topics in ice growth. And the unidirectional freezing pattern stems from perturbations of both basal and edge planes. To date there is no direct comparison of unidirectional freezing behavior between basal and edge plane ice. Here, we in-situ investigate the planar instability as well as the unidirectional freezing pattern of basal and edge planes of ice by a design of parallel freezing samples with specified ice orientations in a NaCl solution as a modeled sea water. The planar instability is discussed via neutral stability curves with surface tension anisotropy for both basal and edge plane ice. For the first time, we realize the simultaneous observation of solid/liquid interfaces of basal and edge plane ice under the same set of freezing conditions. The results show that planar instability occurs faster for edge plane ice than basal plane ice. The time-lapse observations confirm a transient competitive interaction of perturbations between the basal and edge planes ice, which is explained by the anisotropic growth of perturbations in basal and edge planes of ice. These experimental results provide a link between morphology evolution of unidirectional grown sea ice and different ice orientations and are suggested to enrich our understanding of sea ice growth as well as crystallization pattern of other anisotropic materials.

## 1. Introduction

Sea ice growth has attracted a lot of attention by the researchers in the past several decades for its applications in the fields of physical chemistry [1-4], geophysics [5-7] and material science[8]. Pattern formation during sea ice growth is typical of lamellar microstructure of differently oriented ice platelets with brine channels[9-15]. A great number of reports on sea ice growth were based on free growth experiments of polycrystalline ice, and unidirectional growth with carefully oriented ice crystal have been much less focused. The central physical process of sea ice growth originates from the distinct growth behaviors between basal and edge planes of a single ice crystal. And the departures from planar instability give rise to subsequent different nonequilibrium ice morphologies. Thus, a thorough understanding of the interface stability and the subsequent nonequilibrium freezing pattern of basal and edge planes is of great significance for sea ice growth.

---

\* Corresponding author. zhjwang@nwpu.edu.cn
\* Corresponding author. wlilin@nwpu.edu.cn

Harrison and Tiller[16], the pioneer of unidirectional growth of sea ice, investigated the unidirectional freezing pattern of polycrystalline ice in aqueous solutions and reported two types of cells growing in basal plane and C-axis directions, respectively. However, it should be pointed out that sea ice can hardly develop into cellular arrays perpendicular to the basal plane since growth is greatly preferred in directions perpendicular to C-axis, as shown in previous free growth experiments[17-19]. Nagashima and Furukawa[20] reported the nonequilibrium growth of basal plane ice in a confined space between the two separated glass plates in a unidirectional freezing manner, where the cellular growth of basal plane ice was revealed but without the information of the growth of edge plane ice. In fact, the nonequilibrium growth for edge plane ice[21, 22] is distinct from the case for basal plane ice in unidirectional freezing as previously shown by Zhang et al[23].

Planar instability as well as the subsequent freezing patterns for basal and edge plane ice constitute the central scientific problems in sea ice growth, which remains poorly investigated. Up to now, although numerous investigations have been carried out to show the morphology of ice in both free growth[17-19] and unidirectional growth[24] conditions, the relative instability for basal and edge plane ice as well as their subsequent nonequilibrium growth patterns has not been well elucidated yet. Besides, tilted growth of ice is complex, but usually treated simply by a traditional case of tilted growth in unidirectional solidification[25, 26], which requires further experimental clarification.

In this work, by a careful design of two parallel freezing samples with differently oriented ice in a NaCl solution as a modeled sea water, we realize the set of identical freezing conditions between basal and edge plane ice during unidirectional freezing. This experimentally enables us to demonstrate the in-situ simultaneous observations of unidirectional grown basal and edge plane ice in a comparative manner to explore the relative stability of their planar S/L interfaces. By further increment of freezing velocity, we obtain their distinct nonequilibrium growth patterns under the same freezing conditions. The in-situ comparison of unidirectional freezing process between the two distinct ice orientations vividly reflects the effects of ice orientation on the planar instability as well as the subsequent nonequilibrium freezing patterns in sea ice growth. On the basis of these in-situ observations, we discuss the effects of anisotropy of ice in the light of crystal growth theory.

2. Experimental setup

NaCl solution (0.1M) was prepared in the method reported elsewhere [27]. Rectangular glass capillary with an inner space dimension of $0.05 \times 1$ mm$^2$ (VitroCom brand) was utilized for unidirectional freezing experiments. More details on the experimental setup can be found in Ref.[28]. The thermal gradient of the unidirectional freezing platform in this work was previously measured to be linearly distributed on both sides of the S/L interface[29].

The lamellar morphology of unidirectional grown sea ice was previously proposed by Harrison and Tiller[16] as the formation of two types of cells parallel and normal to the C-axis of ice, as shown in **Fig. 1 (a)**. With increment of constitutional undercooling, they concluded that the knife-edged cells would become scallop-edged cells, as shown in **Fig. 1 (b)**. However, scallop-edged cells may be unstable and can hardly be observed under steady state, since increasingly deep grooves among knife-edged cells should reduce the solute pile-up at the S/L interface and retard the increment of constitutional undercooling. In order to reveal the formation process of lamellar morphology in details, we design specific unidirectional freezing experiments by placing two parallel capillaries with distinct well-designed ice orientations, as shown in **Fig. 1 (c)**.

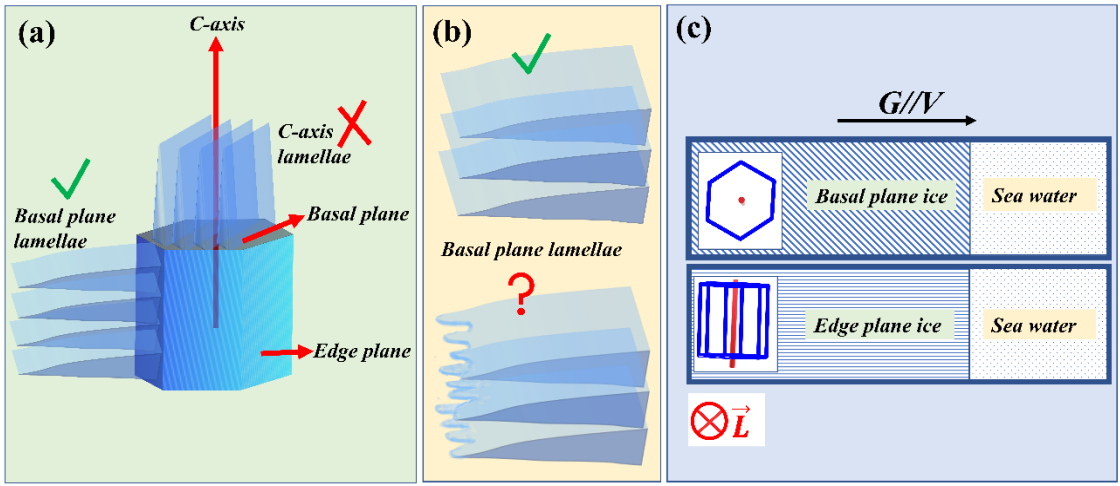

**Fig. 1 (a)** Schematic illustration of basal-plane cells and c-axis cells proposed by Harrison and Tiller[16]; **(b)** Schematic illustration of knife-edged cells (upper) and scallop-edged cells (lower); **(c)** Top view of unidirectional freezing of two parallel capillaries with basal plane ice (represented by shaded area of diagonal lines) and edge plane ice (represented by shaded area of horizontal lines) in contact with liquid phase (represented by shaded area of dots) of sea water, forming two different S/L interfaces. The ice orientations with respect to the unidirectional freezing platform are determined by the relations among thermal gradient $G$ / the freezing velocity $V$, C-axis of ice (represented by red solid rods) and incident light $\vec{L}$ of the microscope (represented by a solid circle with a tilted cross inside in red). The basal plane ice is represented by a hexagon in dark blue solid line with a C-axis in red solid line parallel to $\vec{L}$, as shown in the inset on the left of the upper capillary. The edge plane ice is represented by a rectangle in dark blue solid line with a C-axis in red solid line normal to both $\vec{L}$ and $G \parallel V$, as shown in the inset on the left of the lower capillary.

**Figure 1 (c)** depicts the realization of two parallel freezing samples of basal and edge plane ice in two separate capillaries filled with sea water on a unidirectional freezing platform, in which the freezing conditions are identical except for their ice orientations as labeled on the left of the capillaries. Prior to each freezing experiment, single ice crystal of two specific orientation relations with respect to both the direction of thermal gradient ($G$)/freezing velocity ($V$) and incident light ($\vec{L}$) are prepared. For basal plane ice, the basal plane {0001} of single ice crystal is made nearly parallel to freezing velocity ($V$) / thermal gradient ($G$) that is represented by the solid arrow on the top of **Fig. 1 (c)** but perpendicular to incident light ($\vec{L}$) of microscope that is indicated by a solid circle with a tilted cross inside in red. The basal plane ice in the freezing platform can be represented by a set of relationships of $V(G) \parallel \{0001\}$, $\vec{L} \perp \{0001\}$ and $V(G) \perp \vec{L}$. For edge plane ice, the basal plane $\{0001\}$ of single ice crystal is made nearly parallel to freezing velocity ($V$) / thermal gradient ($G$) that is represented by the solid arrow on the top of **Fig. 1 (c)** and also parallel to incident light ($\vec{L}$) of microscope that is indicated by a solid circle with a tilted cross inside in red. In addition, $V(G)$ is perpendicular to $\vec{L}$. Similarly, the edge plane ice in the freezing platform can be represented by another set of relationships of $V(G) \parallel \{0001\}$, $\vec{L} \parallel \{0001\}$ and $V(G) \perp \vec{L}$.

Single ice crystal with a specific crystal orientation can be kept in a thin glass capillary. By a design of two parallel capillaries with basal and edge plane ice as shown in **Fig. 1 (c)**, we can realize the in-situ comparison experiments among different ice orientations in which the growth conditions are identical for the parallel samples. Planar instability process as well as nonequilibrium freezing patterns of basal and edge plane ice are simultaneously recorded by a charge coupled device (CCD) camera under various freezing velocities.

### 3. Experimental results

Different instants of the S/L interface morphologies for basal and edge plane ice in the NaCl solution under three different freezing velocities $V$ = 5.05 μm/s, 10.03 μm/s, and 38.91 μm/s, are summarized in **Fig. 2-4**, respectively. The steady-state S/L interface morphologies for basal and edge plane ice in **Fig. 2-4** is further summarized in **Fig. 5**. The ice orientations of basal and edge plane ice are indicated by the inset on the left of **Fig. 2-5**. For **Fig. 2-5**, the S/L interface of basal plane ice is in the upper capillary and that of edge plane ice is in the lower capillary in real space.

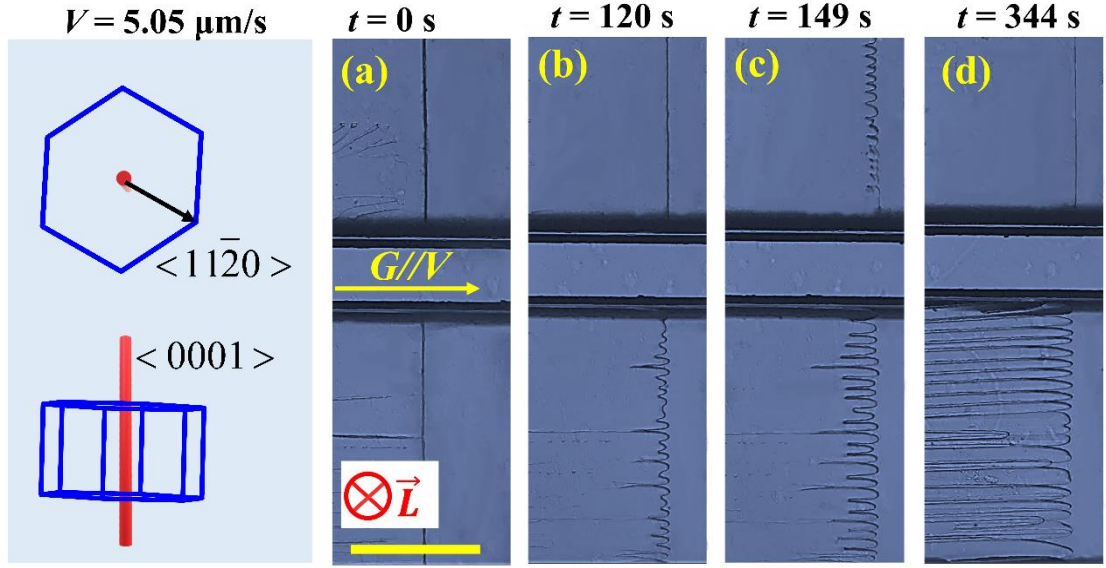

**FIG. 2** Comparison of different instants of S/L interface morphologies of two parallel freezing samples of basal (in the upper capillary) and edge plane ice (in the lower capillary) in a 0.1M NaCl solution under a freezing velocity of 5.05 μm/s, with a thermal gradient of 3.74 K/mm. The inset on the left of the figure represents the ice orientations of the two freezing samples of basal and edge plane ice with labeled crystallographic orientations. The solid arrow in the middle of **(a)** represents the directions of paralleled freezing velocity $V$ and thermal gradient $G$. The direction of incident light ($\vec{L}$) of microscope is indicated by a solid circle with a tilted cross inside in red as shown on the lower left of **(a)**. The scale bar in each figure was 250 μm.

**Figure 2** shows the different instants of the S/L interface morphologies for basal and edge plane ice under a freezing velocity $V$ of 5.05 μm/s. The initial planar S/L interface at static is shown in **Fig. 2 (a)**. And S/L interface at different instants is shown in **Fig. 2 (b-d)**. It can be seen from **Fig. 2 (b)** that planar instability occurs much earlier for edge plane ice than basal plane ice, and the perturbations further develop into cells. The cell tips of edge plane ice gradually become knife-edged from a rounded geometry. For basal plane ice, small perturbations of shallow cells are also observed, which diminish quickly in about 120 s and produce a steady planar S/L interface afterwards.

The different growth of the perturbations for basal and edge plane ice observed in **Fig. 2** indicates a dynamic competitive interaction of perturbations between the two ice orientations. The perturbations in edge plane ice develop faster than basal plane ice, as shown in **Fig. 2 (a-b)**, but solute pile-up still induces the cellular perturbations of basal plane ice to grow via an increasing interface undercooling at the early stages of their growth, as shown in **Fig. 2 (c)**. As time goes by, the perturbations of edge plane ice grow with increasingly deep grooves or brine channels among arrays of cellular perturbations, which provides an increasingly effective segregation space to decrease the solute pile-up at the S/L interface. In fact, the perturbations in thickness direction of basal plane ice resemble the perturbations in edge plane ice and also serve as an increasingly effective segregation space to decrease the solute pile-up at the S/L interface, which is validated by the deceased thickness of basal plane ice via a comparison the S/L interfaces of basal plane ice between **Fig. 2 (a)** and **Fig. 2 (d)**. As a consequence, the interface undercooling is finally reduced and diminishes the perturbations in basal plane ice, which regains a planar S/L interface, as shown in **Fig. 2 (d)**.

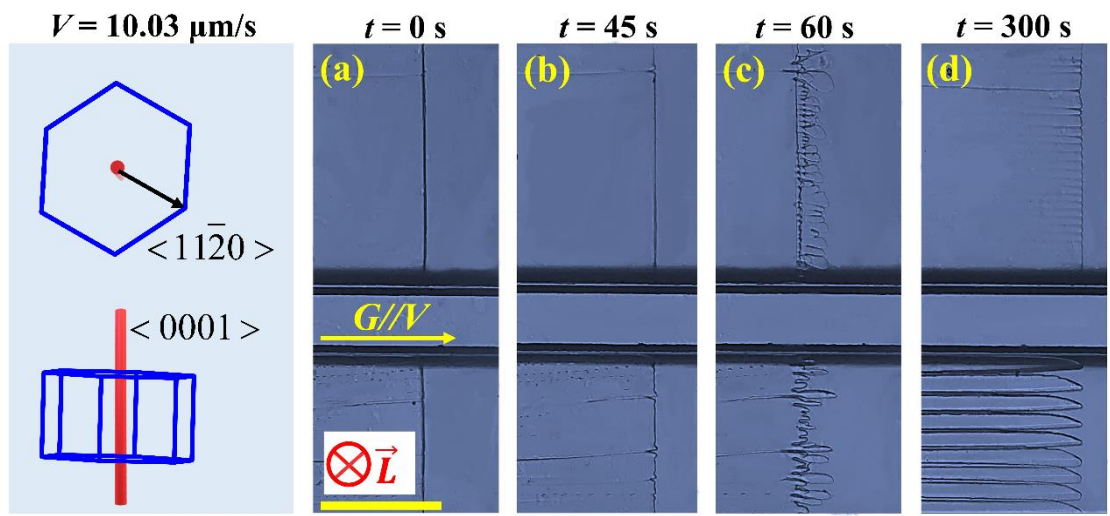

**FIG. 3** Comparison of different instants of S/L interface morphologies of two parallel freezing samples of basal (in the upper capillary) and edge plane ice (in the lower capillary) in a 0.1M NaCl solution under a freezing velocity of 10.03 μm/s, with a thermal gradient of 3.74 K/mm. The inset on the left of the figure represents the ice orientations of the two freezing samples of basal and edge plane ice with labeled crystallographic orientations. The solid arrow in the middle of **(a)** represents the directions of paralleled freezing velocity $V$ and thermal gradient $G$. The direction of incident light ($\vec{L}$) of microscope is indicated by a solid circle with a tilted cross inside in red as shown on the lower left of **(a)**. The scale bar in each figure was 250 μm.

When we increase the freezing velocity $V$ to 10.03 μm/s, a series of morphological changes are obtained at different instants. The details are shown in **Fig. 3**. The initial planar S/L interface at static is shown in **Fig. 3 (a)**. And S/L interface at different instants is shown in **Fig. 3 (b-d)**. It can be seen in **Fig. 3 (b)** that the planar instability of edge plane ice still occurs earlier than basal plane ice. The perturbations of thin ice platelets grow out of the initially planar S/L interface of basal plane ice and enlarge their tips to form a series of ice disks instead of vanishing over time, as shown in **Fig. 3 (c)**. And after a process of competitive growth, a single ice platelet composed of rounded shallow cells parallel to thermal gradient are formed, as shown in **Fig. 3 (d)**. On the contrary, faceted deep cells are observed for edge plane ice.

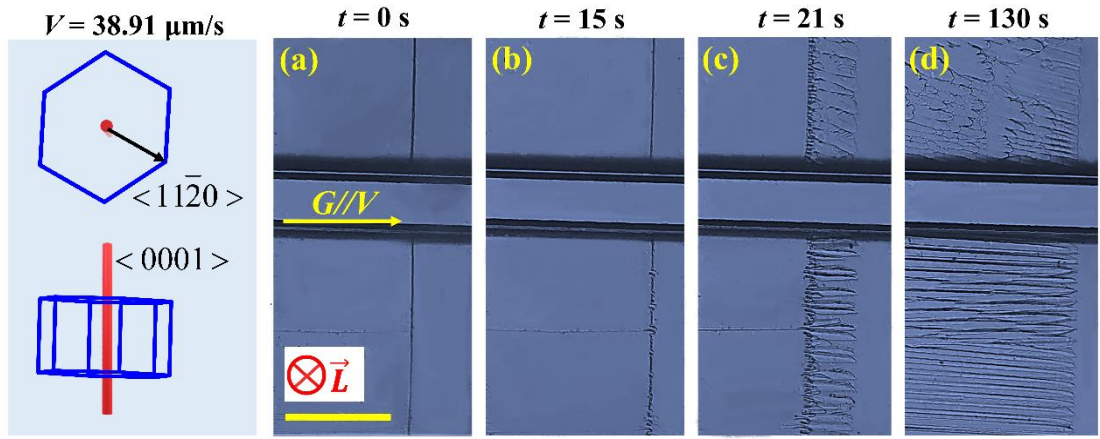

**FIG. 4** Comparison of different instants of S/L interface morphologies of two parallel freezing samples of basal (in the upper capillary) and edge plane ice (in the lower capillary) in a 0.1M NaCl solution under a freezing velocity of 38.91 μm/s, with a thermal gradient of 3.74 K/mm. The inset on the left of the figure represents the ice orientations of the two freezing samples of basal and edge plane ice with labeled crystallographic orientations. The solid arrow in the middle of **(a)** represents the directions of paralleled freezing velocity $V$ and thermal gradient $G$. The direction of incident light ($\vec{L}$) of microscope is indicated by a solid circle with a tilted cross inside in red as shown on the lower left of **(a)**. The scale bar in each figure was 250 μm.

Further increment of freezing velocity $V$ to 38.91 μm/s leads to the observations in **Figure 4**. The initial planar S/L interface at static is shown in **Fig. 4 (a)**. And S/L interface at different instants is shown in **Fig. 4 (b-d)**. It can be seen in **Fig. 4 (b)** that planar instability still occurs earlier for edge plane ice than basal plane ice. On one hand, for basal plane ice, shallow cells are transformed into deep cells with much longer grooves among ice cells, which is in qualitative agreement with the cases in other solidifying materials both experimentally[30, 31] and computationally[32, 33]. In contrast, for edge plane ice, only deep cells are observed in the experiments for all the tested freezing velocities, possibly due to the strong anisotropy in edge plane of ice. On the other hand, **Fig. 4 (d)** shows that for basal plane ice, the effect of unparallel preferred orientation $<11\bar{2}0>$ becomes increasingly predominant and result in tilted growth of ice tips with respect to the direction of thermal gradient, which belongs to a case of classical tilted growth. In contrast, for edge plane ice in **Fig. 4 (d)**, tilted growth also occurs for a preset of paralleled preferred orientation and thermal gradient. Wang et al[34] showed that the actual growth direction or tilt angle $\alpha$ of dendritic tip depends mainly on the spacing Peclet number $P = \lambda_1 V_p / D_L$ ($\lambda_1$ being the primary spacing, $V_p$ being the pulling velocity, and $D_L$ being the solute diffusivity) and misorientation angle $\alpha_0$ with respect to the preferred orientation. For basal plane ice with $\alpha_0 > 0$, the tilted growth can be rationalized by classical tilted growth model. For edge plane ice with $\alpha_0 = 0$, the physical origins of its tilted growth are obviously distinct from the classical cases. The results of **Fig. 2-4** collectively indicate that the transient planar instability behaviors are different for basal and edge plane ice under the same unidirectional freezing conditions.

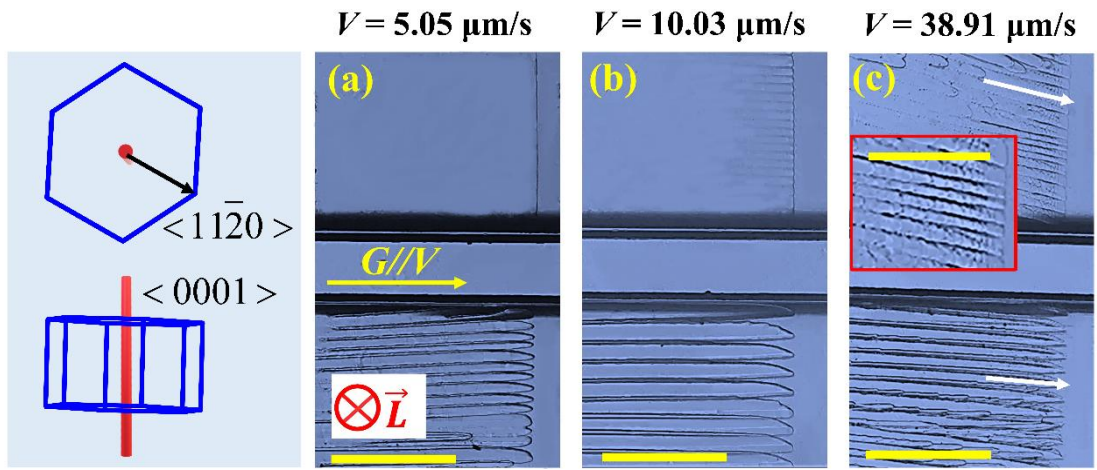

**FIG. 5** Comparison of steady-state S/L interface morphologies of two parallel freezing samples of basal (in the upper capillary) and edge plane ice (in the lower capillary) in a 0.1M NaCl solution under different freezing velocities of **(a)** 5.05 μm/s, **(b)** 10.03 μm/s and **(c)** 38.91 μm/s, respectively with a thermal gradient of 3.74 K/mm. The inset on the left of the figure represents the ice orientations of the two freezing samples of basal and edge plane ice with labeled crystallographic orientations. The solid arrow on the left of the figure represents the directions of paralleled freezing velocity $V$ and thermal gradient $G$. The inset in the middle left of **(c)** is a magnified picture of scalloped lines on the cells of basal plane ice with a scale bar of 125 μm. The white solid arrows in **(c)** indicate the actual growth directions of ice tips for basal and edge plane ice. The direction of incident light ($\vec{L}$) of microscope is indicated by a solid circle with a tilted

cross inside in red as shown on the lower left of **(a)**. The scale bar in each figure was 250 μm.

**Figure 5 (a-c)** represent the steady-state S/L interface morphologies of the basal and edge plane ice under different freezing velocities $V$ (5.05 μm/s, 10.03 μm/s and 38.91 μm/s, respectively) and a given thermal gradient $G$ of 3.74 K/mm, in which their interface undercoolings appear almost identical. **Figure 5** shows that the S/L interface of edge plane ice is more stable than that of basal plane ice under steady state of unidirectional freezing. In **Fig. 5 (a)**, at a low freezing velocity $V$ of 5.05 μm/s, the basal plane ice eventually evolves into a single ice platelet in capillary with a planar S/L interface. In contrary, nearly parallel lamellae with cellular tip are observed in the parallel sample of edge plane ice. As the freezing velocity $V$ increases to 10.03 μm/s, the S/L interface of basal plane ice no longer remains stable under steady state and gradually splits into a series of shallow cells growing parallel to freezing velocity $V$ with a slightly perturbed envelop of S/L interface, while the morphology of the parallel sample of edge plane ice remain lamellar and nearly parallel to freezing velocity $V$ with only some adjustment in its primary spacing as shown in **Fig. 5 (b)**.

In **Fig. 5 (c)** as well as the inset in the middle left of it, the cells in the basal plane ice become rough with many scalloped lines periodically appear behind ice tips in direction perpendicular to the freezing velocity $V$. On the other hand, obvious tilting and overlap of ice lamellae with tilted side branches are observed simultaneously in the parallel sample of edge plane ice, which provides physical clues for the formation of periodically distributed scalloped lines in the basal plane ice. Tilting together with tilted side branches usually occurs at relatively large freezing velocity for edge plane ice. Thus, when tilting is not obvious under low freezing velocity, cellular ice in basal plane ice appears smooth. When tilting becomes significant under high freezing velocity, tilting as well as side branches becomes obvious as indicated by the S/L morphology of the edge plane ice in **Fig. 5 (c)**. And tilting as well as side branches also occur simultaneously in thickness direction of basal plane ice but only to a limited extent since growth in thickness direction is constrained by the thin capillary of basal plane ice.

### 4. Discussions

One of the central scientific questions extracted from our results is the relative stability of planar S/L interface between basal and edge plane ice. From the results of **Fig. 2-4** it can be concluded that planar instability occurs faster for edge plane ice than basal plane ice over time, and formation of cells perpendicular to C-axis instead of along C-axis is physically preferential in the NaCl solution as previously suggested by Harrison and Tiller[16]. Even if perturbations occur in basal plane ice, they are expected to grow perpendicular to C-axis or grow in a tilted manner at larger freezing velocity, and formation of C-axis cells is physically unrealistic.

Our explanations on the experimental results are mainly based on the perspectives of anisotropy properties of ice in the light of crystal growth theory. Energetically speaking, basal plane of ice has a lower surface tension than edge plane of ice[35], and when perturbations develop, the total area of surface with lower surface tension is expected to increase to minimize the total free energy of the system. This may qualitatively explain why the S/L interface of edge plane ice is more stable than that of basal plane ice under steady state of unidirectional freezing, given that breaking down of a basal plane platelet will introduce excessive area of edge plane surface with higher surface tension. It was shown theoretically that the increase of surface tension anisotropy will enlarge the instability of a planar S/L interface[36, 37] and induce faster occurrence of planar instability[38] during unidirectional solidification of a binary alloy. In addition, Wang et al[39, 40] experimentally proved that the effect of surface tension anisotropy on interface stability is significant in studying the planar instability of [111] and [100] orientations in a succinonitrile alloy, in which the S/L interface of [100] orientation with a higher surface tension anisotropy exhibited more unstable interface than [111] orientation.

For the thin samples of basal and edge plane ice in this work, we consider their in-plane anisotropy via two-dimensional surface tension function $\gamma(\theta)$ in the form of "m-fold" rotational symmetry as

$$\gamma(\theta) = \gamma_0[1 + \varepsilon_m \cos(m\theta)] \qquad \text{(Eq. 1)}$$

where $\gamma_0$ is the surface tension in the preferred orientation, $\theta$ is the deviation angle from the preferred orientation, $m$ is the number of rotational symmetry in a given crystal plane, and $\varepsilon_m$ is the anisotropic parameter. Based on previous surface tension data of different crystallographic planes of ice[35], it is easy to obtain the $\gamma(\theta)$ for basal and edge plane ice as

$$\begin{cases} \gamma_{basal}(\theta) = 28.6[1+0.0191\cos 6\theta](mJ/m^2) \\ \gamma_{edge}(\theta) = 27.2[1+0.0485\cos 2\theta](mJ/m^2) \end{cases} \quad \text{(Eq. 2)}$$

Here the $\gamma_{edge}(\theta)$ in Eq. 2 is derived based on an average value of two different edge planes $\{11\bar{2}0\}$ and $\{10\bar{1}0\}$. It should be noted that the derived $\varepsilon_m$ for basal and edge plane ice from their surface tension data deviate largely from the geometrical measurement of globule formed by ice-water equilibrium interface by Koo et al[41], which suggested that basal plane ice had a much lower surface tension anisotropy of 0.001 – 0.003 than edge plane ice with a surface tension anisotropy of 0.3.

By asymptotic analysis[36, 37], the neutral stability curve for planar S/L interface with anisotropic surface tension in plane $(\beta, v)$ are determined by the nonlinear equations as

$$\begin{cases} \beta = v[1 - v\hat{a}_c^2(1-(m^2-1)\varepsilon_m) - (2vk)^{1/2}(\hat{a}_c^2+1/4)^{1/4}] \\ k = 2v(\hat{a}_c^2+1/4)^{1/2}[(\hat{a}_c^2+1/4)^{1/2}+k-1/2]^2 \end{cases} \quad \text{(Eq. 3)}$$

where $\beta = \dfrac{\gamma_0 T_m G}{[-m_L C_\infty (1-k)/k]^2 \Delta H}$ is dimensionless thermal gradient, $v = \dfrac{\gamma_0 T_m V}{m_L C_\infty (1-k) D_L \Delta H / k}$ is the dimensionless freezing velocity, $\hat{a}_c$ is the dimensionless critical wave number, $k$ is the partition coefficient of solute in ice, $\Delta H$ is the latent heat per unit volume due to phase change, $m_L$ is the liquidus slope for water-NaCl system[29], $C_\infty$ is the solute concentration far from the S/L interface, $D_L$ is the solute diffusivity in the NaCl solution[29], and $T_m$ is the melting point of ice in pure water. By utilizing the values for the physical parameters in **Tab. 1**, we plot the neutral stability curve for basal and edge plane ice based on the surface tension functions in Eq. 2 in combination with the numerical solutions of Eq. 3, as shown in **Fig. 6**. In addition, all data points from our experiments are transformed into dimensionless form in plane $(\beta, v)$ based on the physical parameters in **Tab. 1** and also plotted in **Fig. 6** for comparison.

**TAB. 1** Physical parameters utilized for numerical solutions of Eq. 3 as well as the dimensionless transformation of data points from our experiments.

| Parameters | Value | Units (SI) |
|---|---|---|
| $k$ | 0.0001 | 1 |
| $T_m$ | 273.15 | K |
| $\Delta H$ | 3.06 | $10^8$ J/m$^3$ |
| $D_L$ | 7.74[29] | $10^{-10}$ m$^2$/s |
| $-m_L C_\infty$ | 0.356[29] | K |
| $G$ | 3.74 | $10^3$ K/m |
| $V$ | 5.05, 10.03, 38.91 | $10^{-6}$ m/s |

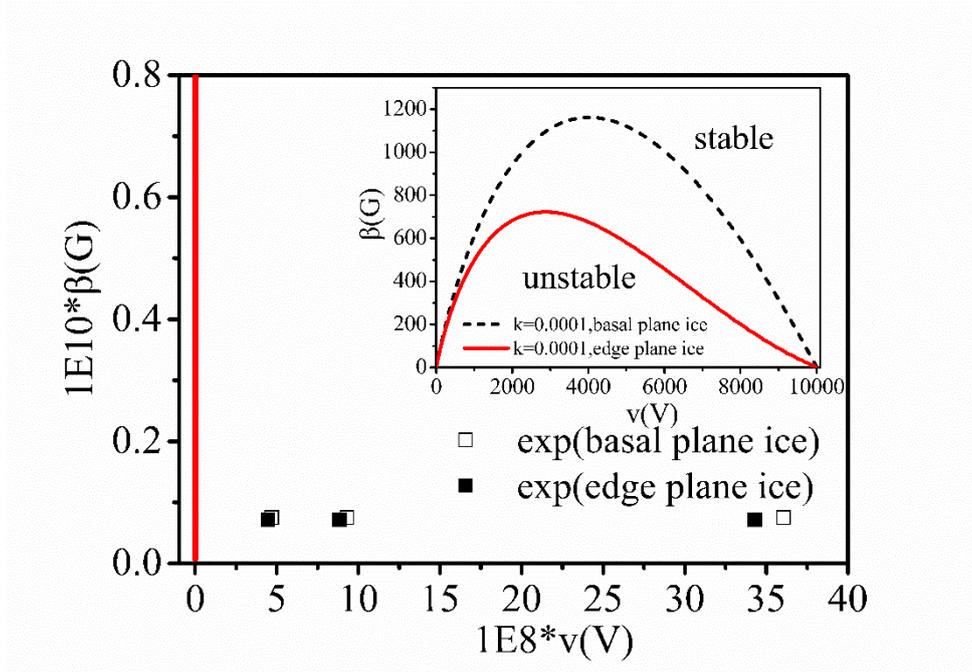

**FIG. 6** Comparison between calculated neutral stability curves for basal/edge plane ice via Eq. 6 and dimensionless data points from the experiments in a 0.1M NaCl solution under different freezing velocities of 5.05 μm/s, 10.03 μm/s and 38.91 μm/s with a thermal gradient of 3.74 K/mm. The transformed data points from the experiments are found to be very close to the original point. The inset figure represents the global shape of the neutral stability curves for basal and edge plane ice in plane $(\beta, v)$.

Since the dimensionless data points from our experiments are very close to the original point, the obtained neutral stability curves in **Fig. 6** are magnified close to the original point of plane $(\beta, v)$. And the global shape of the neutral stability curves for basal and edge plane ice in plane $(\beta, v)$ is given in the inset figure of **Fig. 6**. It can be seen from the inset of **Fig. 6** that the neutral stability curve for basal plane ice always covers a much wider region than edge plane ice, indicating that basal plane ice is more unstable than edge plane ice against all perturbations. However, our experimental freezing conditions are very close to the original point of plane $(\beta, v)$. Thus, we focus on the magnified region of neutral stability curves near the original point. From **Fig. 6**, it can be seen that the two neutral stability curves almost overlap near the original point, and all data points from our experiments fall into the unstable region below the neutral stability curves, which is in agreement with the observed planar instability for both basal and edge plane ice as shown in **Fig. 2-4**. In fact, the observed steady-state planar interface of basal plane ice in **Fig. 2 (d)** is not a result of a stable planar S/L interface but a result of dynamic competition of perturbations in basal and edge plane of ice after their planar instability. This indicates that the consideration of surface tension anisotropy can not be applied to address the observed steady-state morphologies between basal and edge plane ice. Therefore, it is suggested that lamellar morphology with smooth ice platelets in sea ice must be addressed based on the dynamic interactions among perturbations of shallow cells in basal and edge plane of ice after their planar instability, in which the system enters nonlinear regime and kinetic anisotropy of ice is suggested to play a significant role.

Unfortunately, up to now few investigations considered the kinetic anisotropy in addressing the planar instability of an anisotropic system. And dynamic interaction among shallow cells is notoriously complicated. Future investigations are needed which take the kinetic anisotropy of ice into consideration so as to explain the different transient behaviors of planar instability as well as the dynamic interactions among perturbations for basal and edge plane ice after their planar instability.

Another essential scientific question arising from our results is the two distinct types of tilted growth for basal and edge plane ice as validated by our in-situ experiments. Tilted growth of unidirectional grown ice lamellae is often interpreted in the context of a combined effect of a misaligned orientation of surface-tension anisotropy with thermal flux[25, 42], which remains skeptical about its physical basis in the case of edge plane ice. In the freezing sample of basal plane ice, tilting occurs at increased freezing velocity from initially shallow cells parallel to the thermal flux as shown in **Fig. 4 (b)**. Further increment of freezing velocity results in tilting of ice tips with respect to thermal flux as shown in **Fig. 4 (c)** and the tilting direction gradually coincides with the preferred crystallographic orientation $<11\bar{2}0>$. This belongs to a classical case of tilted growth in unidirectional solidification, in which a combined effect of unparalleled preferred orientation $<11\bar{2}0>$ and thermal gradient is considered. In this case, tilting direction of ice tips is suggested to be restricted to the direction between preferred orientation and thermal gradient. On the contrary, a distinct case of tilted growth is observed in the parallel freezing sample of edge plane ice. The preferred orientation of edge plane ice is set almost parallel to thermal flux, yet continuous tilting of ice tips still occurs upon increased freezing velocity, which is unconstrained by the direction between preferred orientation and thermal flux and can not be addressed by classical theory of tilted growth in unidirectional solidification[23, 29].

## 5. Conclusion

In conclusion, our in-situ experiments reveal the formation mechanism of lamellar sea ice by providing a direct comparison of the relative interface stability between basal and edge plane and the edge plane of ice under the set of identical unidirectional freezing conditions. The transient processes of planar instability between basal and edge plane ice are observed. For the first time, we observed a transient competitive interaction of perturbations between basal and edge plane ice, which stabilizes the perturbed S/L interface of basal plane ice and makes it regain a planar S/L interface. Based on the neutral stability analysis which couples surface tension anisotropy of basal and edge plane ice, it is found that the data points of our experiments fall into an unstable region for both basal and edge plane ice. However, this method can not be applied to explain the transient behaviors of planar instability as well as the formation of steady-state lamellar morphology during the unidirectional freezing experiments. Kinetic anisotropy of ice is suggested to be a key factor in the observed transient planar instability behaviors as well as the steady-state S/L interface morphologies between basal and edge plane ice. In addition, tilting behavior of ice tips is observed for both basal and edge plane ice in the NaCl solution under the same freezing conditions, but their physical origins are distinct. Collectively, these observations provide a link between morphology evolution of unidirectional grown sea ice and different ice orientations and are suggested to enrich our understanding of sea ice growth as well as nonequilibrium crystallization pattern of other anisotropic materials.

**Acknowledgements**

The work was supported by the Research Fund of the State Key Laboratory of Solidification Processing (NPU), China (Grant No. 2020-TS-06, 2021-TS-02).